\documentclass[11pt,a4paper]{article}
\usepackage[margin=2.5cm]{geometry}
\usepackage{amsmath,amssymb}
\usepackage{graphicx}
\usepackage{bbold}
\usepackage{tikz}
\usepackage{xcolor}

\definecolor{midsurf}{RGB}{244,238,222}\definecolor{gridc}{RGB}{160,152,135}
\definecolor{faceblue}{RGB}{60,110,200}\definecolor{facelat}{RGB}{235,140,40}
\newcommand{\Zt}{\zeta}   % transverse coordinate (was xi)
\newcommand{\Vp}{\varpi}   % Frank energy density (was w)
\usetikzlibrary{arrows.meta}
\newcommand{\junction}[2]{%
\begin{tikzpicture}[x=1cm,y=1cm,>={Stealth[length=2mm]}]
  \def\W{4.6}\def\H{2.6}
  \fill[blue!8] (0,0) rectangle (\W,\H);
  \draw[line width=0.5pt] (0,0) -- (\W,0);\draw[line width=0.5pt] (0,\H) -- (\W,\H);
  \draw[magenta!80!black,line width=1.1pt] (0,0) -- (0,\H);
  \draw[magenta!80!black,line width=1.1pt] (\W,0) -- (\W,\H);
  \node[magenta!80!black,font=\small,below] at (0,0) {$v=0$};
  \node[magenta!80!black,font=\small,below] at (\W,0) {$v=2\pi$};
  \node[font=\scriptsize,anchor=west] at (\W+0.15,\H) {$u=+\tfrac w2$};
  \node[font=\scriptsize,anchor=west] at (\W+0.15,0) {$u=-\tfrac w2$};
  \draw[->,green!50!black,line width=0.9pt] (0.55,0.3) -- (0.55,0.95) node[above,font=\scriptsize] {$\mathbf e_u$};
  \draw[->,orange!80!black,line width=0.9pt] (0.55,0.3) -- (1.2,0.3) node[right,font=\scriptsize] {$\mathbf e_v$};
  \foreach \y/\aL in {2.05/62, 1.3/36, 0.55/10}{
     \pgfmathsetmacro{\dx}{0.38*sin(\aL)}\pgfmathsetmacro{\dy}{0.38*cos(\aL)}
     \draw[->,red!75!black,line width=1.3pt] (-\dx,\y-\dy) -- (\dx,\y+\dy);
  }
  \ifnum#1=1
    \foreach \yR/\aL/\yL in {0.55/62/2.05, 1.3/36/1.3, 2.05/10/0.55}{
       \pgfmathsetmacro{\aR}{-\aL+180*#2}
       \pgfmathsetmacro{\dx}{0.38*sin(\aR)}\pgfmathsetmacro{\dy}{0.38*cos(\aR)}
       \draw[->,red!75!black,line width=1.3pt] (\W-\dx,\yR-\dy) -- (\W+\dx,\yR+\dy);
       \draw[gray,dash pattern=on 2pt off 2pt,line width=0.5pt] (0,\yL) to[out=0,in=180] (\W,\yR);
    }
  \else
    \foreach \yR/\aL in {2.05/62, 1.3/36, 0.55/10}{
       \pgfmathsetmacro{\aR}{\aL-180*#2}
       \pgfmathsetmacro{\dx}{0.38*sin(\aR)}\pgfmathsetmacro{\dy}{0.38*cos(\aR)}
       \draw[->,red!75!black,line width=1.3pt] (\W-\dx,\yR-\dy) -- (\W+\dx,\yR+\dy);
       \draw[gray,dash pattern=on 2pt off 2pt,line width=0.5pt] (0,\yR) -- (\W,\yR);
    }
  \fi
\end{tikzpicture}}

\newcommand{\be}{\begin{equation}}
\newcommand{\ee}{\end{equation}}

\def\H{H}

\def\ev{{\bf e}}

\def\nv{{\bf n}}
\def\Nv{{\bf N}}

\def\tv{{\bf t}}

\def\Lv{{\bf L}}

\def\Pv{{\bf P}}
\def\Qv{{\bf Q}}

\def\dd{{\rm d}}

\def\grad{\nabla}

\def\grads{\nabla\hspace{-1mm}_s}

\def\dv{{\rm div}}

\newcommand{\ot}{\otimes}

\newcommand{\bnu}{ \mbox{ \hspace{-1mm}\boldmath $\nu$ \hspace{-1mm}}}

\begin{document}
\title{Nematic equilibria on a M\"obius strip: topology and extrinsic curvature}
\author{Massimo Frittelli$^{1}$, S\'ebastien Fumeron$^{2}$, and Gaetano Napoli$^{3}$\thanks{gaetano.napoli@unina.it}\\[2mm]
\parbox{0.92\textwidth}{\centering\small
$^{1}$Dipartimento di Matematica e Fisica ``E. De Giorgi'', Universit\`a del Salento, Lecce, Italy\\[1mm]
$^{2}$Laboratoire de Physique et Chimie Th\'eoriques, UMR Universit\'e de Lorraine--CNRS 7019, 54000 Nancy, France\\[1mm]
$^{3}$Dipartimento di Matematica e Applicazioni ``R. Caccioppoli'', Universit\`a degli Studi di Napoli ``Federico II'', Napoli, Italy}}
\date{}
\maketitle

\begin{abstract}
We investigate the equilibrium configurations of a nematic liquid crystal confined in a thin tubular neighbourhood of a M\"obius strip. Assuming degenerate tangential anchoring and a director uniform across the thickness, we derive an effective two-dimensional Frank energy that accounts for both the intrinsic and extrinsic curvature of the mid-surface. The non-orientability of the substrate imposes identification conditions that couple the reversal of the local frame to the head-to-tail symmetry of the nematic director. We show that the resulting topological classification depends on the parity of the number of half-twists: the sector index is a winding number for even parity, whereas only its parity is physically meaningful for odd parity, in which case the net director rotation is not quantised. We analyse the narrow-strip limit and solve the full two-dimensional equilibrium problem by finite-element methods. The competition between topological sectors depends on the strip aspect ratio and elastic anisotropy. In particular, for a single half-twist, the preferred sector changes at an aspect ratio of approximately 0.58, while for two half-twists two distinct minima of the same sector exchange their energy ordering near 0.60, without any change of topology. These results reveal how non-orientable topology and extrinsic curvature jointly shape nematic equilibria on curved substrates.
\end{abstract}

\section{Introduction}
The geometry of a curved substrate affects orientationally ordered phases in two distinct ways: through its topology, which constrains the defects of the order parameter, and through its curvature, which couples to the order parameter itself \cite{K02,BG09}; for an introduction to the topological defects of nematics see Ref.~\cite{FB23}. In the case of nematic films, the Frank energy must be supplemented by terms that account for the extrinsic geometry of the substrate: in the theory of Napoli and Vergori \cite{NV12,NV12b} the elastic energy of a director field tangent to a surface depends on both the intrinsic and the extrinsic curvature of the surface. Analogous extrinsic couplings are known for columnar and lamellar phases \cite{SVKN07,KNSV09} and for tilted molecules in fluid membranes \cite{JHPP07}. The equilibria of nematic films have been analysed, among others, for the cylinder \cite{NV12}, including the defects of a nematic patch on it \cite{P22}, and for the torus \cite{SSV14}. The effect of the extrinsic curvature on the spontaneous flow of active nematics has been studied on a cylinder in \cite{NT20}. Curvature also controls the distribution and the dynamics of defects: in active nematic toroids defect pairs unbind and segregate in regions of opposite Gaussian curvature \cite{E18}. The mathematical framework is provided by the variational theory of liquid crystals \cite{V94} and by the classical differential geometry of surfaces \cite{dC76}.

Domain orientability is a topological property whose effect on orientationally ordered systems is much less explored than that of the genus or of the Euler characteristic \cite{S26}. On a non-orientable domain a global orientation cannot be defined, so that there is no notion of an arbitrary homogeneous rotation of the order parameter, and the orientational degrees of freedom are subject to constraints that have no counterpart on a sphere or on a torus \cite{S26}. For a nematic the situation is more delicate, since the director is defined only up to $\nv\to-\nv$: the continuity conditions must account for both the reversal of the normal and the head-to-tail symmetry, and, as we show, they depend on the parity of the number of half-twists. Since every non-orientable surface contains a M\"obius strip \cite{MA13}, a nematic confined in a thin neighbourhood of a M\"obius strip, described through its mid-surface, is the elementary instance of the problem.

Non-orientable substrates have received attention in several, rather different, settings, which we recall in order to delimit the present contribution. In the physics of liquid crystals, a Möbius strip has been considered as a colloidal inclusion in a chiral nematic bulk, where its non-orientability enforces the presence of disclination loops and of linked defect structures in the surrounding medium \cite{MA13}; Möbius-strip topology also characterises the localised twist-wall structures (möbiusons) of chiral nematics \cite{ZTWS23}, within the broader class of knotted and solitonic textures reviewed in \cite{S20}; in these works the director is a three-dimensional field in a chiral bulk, and the strip is either a colloidal boundary condition \cite{MA13} or a structure that emerges from the field itself \cite{ZTWS23}. In statistical models of orientational order, the topological constraint has instead been recast as a non-local symmetry relation on the orientable double cover: this is the approach followed in \cite{S26} for XY-type, Vicsek and active Ising models defined on the Möbius strip and on the Klein bottle. A double-cover formulation is also adopted in the study of magnetic textures (bimerons) on a Möbius strip \cite{SJ25}. Finally, the ferromagnetic Möbius ring has been shown to exhibit a coupling between the chirality of the ring and that of the magnetisation texture, which originates from curvature- and torsion-induced terms of Dzyaloshinskii--Moriya type that are linear in the extrinsic curvature \cite{PK15}. These contributions deal with order parameters of a vectorial or magnetic nature, or with the bulk nematic around a strip; none of them concerns a nematic confined in a thin neighbourhood of a Möbius strip, with a director constrained to be tangent to the mid-surface.

To the best of our knowledge, and within the limits of the literature we have been able to examine, the equilibrium problem for a \emph{nematic confined in a tubular neighbourhood of a Möbius strip}, reduced to a director tangent to the mid-surface, with a Frank energy that retains the extrinsic curvature of the embedding, and with the identification conditions that follow from the non-orientability and from the head-to-tail symmetry, has not been addressed in this form. In particular, the dependence of the topological index on the parity of the number of half-twists, and the competition between the corresponding sectors as a function of the geometry and of the elastic anisotropy, appear to be new.

We stress the setting of the present work. The strip is thin and the director is tangent to it (degenerate tangential anchoring on the faces of the ring), so that the only unknown is the angle $\alpha(u,v)$ between the director and the generatrices of the strip, and the problem is a two-dimensional one on the strip itself. The non-orientability does not enter through a three-dimensional bulk director, as for a Möbius colloid in a nematic bulk, nor through the orientable double cover of the strip, as in the statistical models mentioned above: it is encoded in the identification conditions satisfied by $\alpha$ on a fundamental domain of the strip (Sect.~III), together with the head-to-tail symmetry of the director. The problem is not trivial because three ingredients act together: the non-orientable topology of the strip, which fixes the identification conditions and the sectors; the head-to-tail symmetry of the nematic, which makes only the parity of the index physical for $k$ odd; and the extrinsic curvature, which couples the director to the embedding of the strip and makes the equilibria non-uniform and dependent on the aspect ratio.

The paper is organised as follows. We first derive the Frank energy and the associated Euler--Lagrange equation in the parametrisation of the strip (Sect.~II) and we formulate the boundary conditions and the continuity conditions at the identification edge, whose structure depends on the parity of $k$ (Sect.~III). We then study the narrow-strip limit, in which the problem reduces to an ordinary boundary value problem (Sect.~IV). The equilibrium configurations of the complete two-dimensional problem, namely the selected sector as a function of $k$, of the aspect ratio of the strip and of the Frank constants, are described in Sect.~V; the numerical methods and their validation are collected in the appendices. Sect.~VI contains the concluding remarks.

\section{Free energy}

\subsection{Parametrisation of the domain}
We consider a nematic liquid crystal confined in a tubular neighbourhood of a Möbius strip, namely in the three-dimensional Möbius ring of thickness $2h$ described by the spatial domain
\begin{equation}
P(u, v, \xi) =  P_\Sigma(u, v) + \xi \, \bnu, \quad -h \leq \xi \leq h,
\end{equation}
where $ \xi $ is the transverse coordinate, representing the points across the thickness of the ring, which is therefore equal to $2h$. The mid-surface $\Sigma$, i.e.\ the two-dimensional Möbius strip, can be parametrised in the form
\begin{subequations}
\be
x(u,v) = \left(r + u \sin \frac{k v}{2} \right) \cos v
\ee
\be
y(u,v) =\left(r + u \sin \frac{k v}{2}\right) \sin v
\ee
\be
z(u,v) = u \cos \frac{k v}{2}
\ee
where $r>0$ is the radius of the central axis, $v\in [0, 2\pi)$ describes the rotation angle of the generatrices around the central axis, and $u\in[-w/2,w/2] $ describes the position of a point along the generatrices. The parameter $k$ is an integer that measures the number of half-winds of the tape. In the special case where $k=0$, we obtain a cylinder of radius $r$ and height $w$.
\end{subequations}
Throughout we assume $w<2r$, so that $r+u\sin(kv/2)>0$ on the whole domain; we denote by $\varepsilon:=w/(2r)\in(0,1)$ the aspect ratio of the strip.
\begin{figure}[!htbp]
\centering
\resizebox{0.6\textwidth}{!}{\begin{tikzpicture}[x=3.9cm,y=3.9cm,>={Stealth[length=2mm]}]
\input{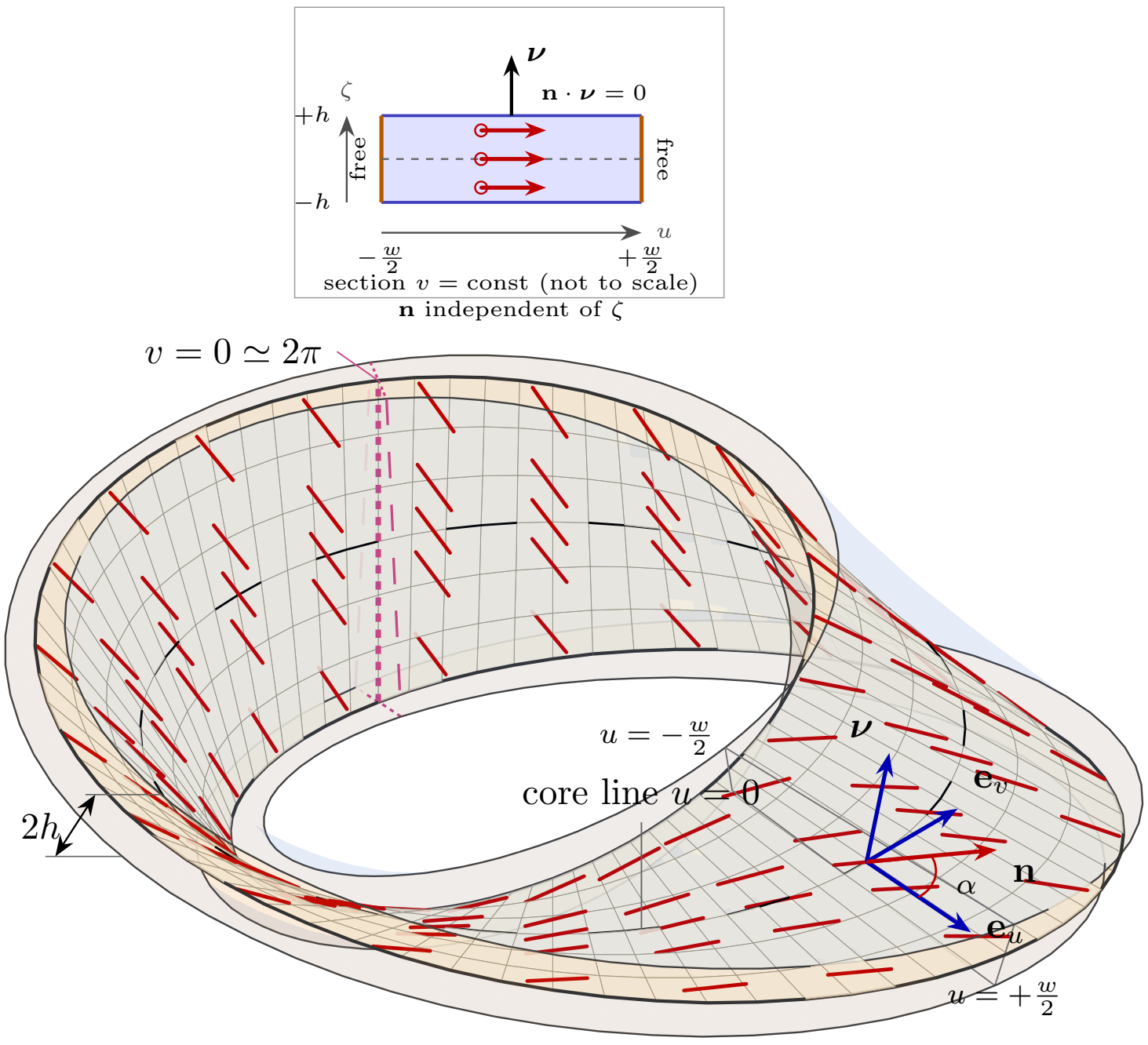}
\end{tikzpicture}}
\caption{Parametrisation of the Möbius ring for $k=1$ ($r=1$, $w=0.9$; the thickness $2h$ is exaggerated). The mid-surface $\Sigma$ (the strip) is drawn with its coordinate lines, and the three-dimensional domain is the transparent slab of thickness $2h$ around it, with its edges outlined and its cross-sections drawn at the cut and at the point of the triad: it is bounded by the faces $\Zt=\pm h$ (blue), which form a single surface since $(u,\Zt)\sim(-u,-\Zt)$ at the cut, and by the free lateral faces $u=\pm w/2$ (orange). The dashed curve is the core line $u=0$, parametrised by $v\in[0,2\pi)$; $u\in[-w/2,w/2]$ is the arc length along the generatrices $v=\mathrm{const}$, and the edges of the strip are $u=\pm w/2$. At the point marked, the orthonormal triad $\{\mathbf e_u,\mathbf e_v,\boldsymbol\nu\}$ is shown (blue), together with the director $\mathbf n=\cos\alpha\,\mathbf e_u+\sin\alpha\,\mathbf e_v$ and the angle $\alpha$. The dashed magenta segment is the identification edge $v=0\simeq2\pi$, across which $(u,0)$ is glued to $((-1)^ku,2\pi)$. The red segments (drawn without heads, as the director is a line field) show the minimiser of the reduced functional of Sect.~IV for $k=1$ in the sector $p$ odd, on the core line and on four parallel lines at $u=\pm0.28\,w/2$ and $u=\pm0.8\,w/2$ (the reduced solution does not depend on $u$). Inset: section $v=\mathrm{const}$ of the slab; the dashed line is $\Sigma$ ($\Zt=0$), the director is tangent to $\Sigma$ ($\mathbf n\cdot\boldsymbol\nu=0$) and independent of $\Zt$, and its component along $\mathbf e_v$ points out of the plane of the section ($\odot$).}
\label{fig:mob}
\end{figure}

\subsection{Local frame}
On the strip we use the orthonormal frame adapted to the parametrisation: $\ev_u$ is the unit tangent to the generatrices $v=\mathrm{const}$, $\ev_v$ is the unit tangent to the curves $u=\mathrm{const}$ and $\bnu$ is the unit normal along which the ring is thickened. In terms of the fixed Cartesian basis $\{\ev_x,\ev_y,\ev_z\}$, the first two vectors are given by
\be
\ev_u = \cos v \sin \frac{k v}{2} \ev_x + \sin v \sin \frac{k v}{2} \ev_y + \cos \frac{k v}{2}\ev_z
\ee
\begin{align}
f_ v \ev_v = \left[k u \cos v \cos \frac{k v}{2} - 2 \sin v \left(r + u \sin \frac{k v}{2} \right)\right]\ev_x \nonumber \\+
\left[k u \sin v \cos \frac{k v}{2} + 2 \cos v \left(r + u \sin \frac{k v}{2} \right)\right]\ev_y \nonumber \\ - k u \sin \frac{k v}{2} \ev_z
\end{align}
where
\[
f_v := \sqrt{k^2 u^2 + 4\left(r + u \sin \frac{k v}{2} \right)^2 }
\]
and the unit normal is defined by $\bnu:=\ev_u\times\ev_v$. The triad $\{\ev_u,\ev_v,\bnu\}$ is orthonormal, and the metric induced on the strip is
\be
\dd s^2=\dd u^2+\frac{f_v^2}{4}\,\dd v^2 ,
\label{metric}
\ee
so that $u$ is the arc length along the generatrices, while $f_v/2$ is the scale factor of the coordinate $v$. In what follows $f_{v,v}$ denotes the derivative of the function $f_v$ with respect to $v$.

\subsection{Curvature tensor}
The way in which the strip is embedded in space is encoded in the curvature tensor $\Lv:=-\grad\bnu$, which is symmetric and tangent to the strip; its components in the frame $\{\ev_u,\ev_v\}$ follow from the derivatives of $\bnu$ along $u$ and $v$ and read
\be
\Lv := -\grad \bnu = -\frac{2 k r}{f_v^2} (\ev_u \ot \ev_v + \ev_v \ot \ev_u) + c_v \ev_v \ot \ev_v
\ee
where
\[
c_v := \frac{4}{{f_v^3}}\left[k^2 u^2 + 2\left(r + u \sin \frac{k v}{2} \right)^2\right] \cos \frac{k v}{2}.
\]
In the frame $\{\ev_u,\ev_v\}$ the components of $\Lv$ are therefore $L_{uu}=0$, $L_{uv}=-2kr/f_v^2$ and $L_{vv}=c_v$. Consequently, recalling that the Gaussian curvature is the determinant of the shape operator \cite{dC76}, we have
\be
K_G=\det\Lv=-\frac{4k^2r^2}{f_v^4}\le 0 ,
\ee
which vanishes only for $k=0$, and the principal curvatures are
\be
\lambda_\pm=\frac12\left(c_v\pm\sqrt{c_v^2+\frac{16k^2r^2}{f_v^4}}\right),
\label{lambda}
\ee
with $\lambda_+\lambda_-=K_G$.

\subsection{Frank free energy}
Let us parametrise the director field $\nv$ as $\nv = \cos \alpha \ev_u + \sin \alpha \ev_v$. Following \cite{NV12,NV12b} (the variational theory of the Frank energy is exposed in \cite{V94}), the Frank energy density is given by
\be
2 \Vp = k_1  \kappa_t^2 + k_2 \tau_n^2 + k_3 (\kappa_n^2 + c_n^2).
\ee
Accordingly, it is necessary to compute the geodesic curvatures $\kappa_n := \tv \cdot (\grads \nv) \nv$ and $\kappa_t := \tv \cdot (\grads \nv) \tv $, the geodesic torsion $\tau_n := - \tv \cdot \Lv \nv $ and the normal curvature $c_n:= \nv \cdot \Lv \nv $, where $\tv:=\bnu\times\nv$ is the unit tangent vector orthogonal to $\nv$. Here $\kappa_t=\dv_s\nv$ measures the splay of the director, $\kappa_n$ is the geodesic curvature of its flux lines (intrinsic bend), $\tau_n$ is the extrinsic twist and $c_n$ the extrinsic bend:
\[
\kappa_n = \cos \alpha \; \alpha_{,u} + \frac{2}{f_v} \sin \alpha \; \alpha_{,v} + \frac{\kappa_v}{f_v^2} \sin \alpha
\]
\[
\kappa_t =  -\sin \alpha \; \alpha_{,u} + \frac{2}{f_v} \cos \alpha \; \alpha_{,v} + \frac{\kappa_v}{f_v^2} \cos \alpha
\]
\[
c_n =  \sin \alpha \left(c_v \sin \alpha - \frac{4}{f_v^2} k r \cos \alpha \right)
\]
\[
\tau_n =  -\frac{c_v}{2} \sin 2 \alpha + \frac{2}{f_v^2} k r \cos 2 \alpha
\]
where
\[
\kappa_v = k^2 u + 4 r \sin \frac{k v}{2} + 4 u \sin^2 \frac{k v}{2}.
\]
We note that $\kappa_v/f_v^2=\partial_u\ln(f_v/2)$ is the geodesic curvature of the curves $u=\mathrm{const}$, and that the spin connection of the metric \eqref{metric} is determined by it.

\paragraph{Hypotheses of the reduction.} The two-dimensional energy density above is obtained from the three-dimensional Frank energy under the following assumptions: (i) the director is tangent to the strip, $\nv\cdot\bnu=0$, as enforced by a degenerate tangential anchoring on the faces $\Zt=\pm h$; (ii) $\nv$ does not depend on $\Zt$; (iii) the thickness is small compared to the minimum radius of curvature $\ell$ of the strip, $h\ll\ell$, where $\ell$ is defined in \eqref{ell} below. Under these hypotheses the elastic energy of the film is $\mathcal F=2h\int_\Sigma \Vp\,\dd A$, up to corrections of relative order $h^2/\ell^2$ \cite{NV12,NV12b}: for a director independent of $\zeta$ the energy per unit area of the surface parallel at distance $\zeta$ is a smooth function of $\zeta$, so that its mean over the symmetric interval $[-h,h]$ differs from its value at $\zeta=0$ by terms even in $h$ (the terms linear in $\zeta$, including that of the volume element, cancel), of relative order $h^2/\ell^2$ for fields varying on the scale $\ell$; we checked numerically the $h^2$ scaling of this difference against the three-dimensional Frank density, for unequal Frank constants. The statement concerns the energy of the ansatz, not its distance from the minimiser of the three-dimensional problem. The case $\nv\cdot\bnu\neq0$, which allows for escape of the director along the thickness and for the removal of defect cores, is not considered, nor is the saddle-splay contribution.

\paragraph{The minimum radius of curvature.} The curvature scale entering hypothesis (iii) is the inverse of the largest principal curvature in modulus attained on the strip. With $\lambda_\pm$ given in \eqref{lambda} we set
\be
\ell(k,r,w):=\left[\max_{(u,v)}\,\max\{|\lambda_+|,|\lambda_-|\}\right]^{-1}.
\label{ell}
\ee
For $w\to0$ the maximum is attained on the core line $u=0$, where $f_v=2r$ and $c_v=\cos(kv/2)/r$, so that $\lambda_\pm=\tfrac1{2r}\big(\cos\tfrac{kv}2\pm\sqrt{\cos^2\tfrac{kv}2+k^2}\big)$; since $|\lambda_\pm|$ increases with $|c_v|$, the maximum is reached for $\cos(kv/2)=\pm1$ and
\be
\ell=\frac{2r}{1+\sqrt{1+k^2}}+O(w),
\label{ell0}
\ee
which for $k=0$ gives the radius $r$ of the cylinder. Equation \eqref{ell0} shows that $\ell$ is controlled by $r$ and $k$ alone at leading order; the dependence on $w$ enters only through the $O(w)$ correction. We computed $\ell$ from \eqref{ell} on the grid of the numerical solution of Sect.~V; the values of $\ell/r$ are collected in Table~\ref{tab:ell}. Hypothesis (iii) therefore requires $h\ll 2r/(1+\sqrt{1+k^2})$, which becomes increasingly restrictive as the number of half-twists grows, and the decrease of $\ell$ with $\varepsilon$ is appreciable only for $k=1,2$.
\begin{table}[h]
\centering
\begin{tabular}{c|ccccc}
$k$ & $\varepsilon\to0$, \eqref{ell0} & $\varepsilon=0.05$ & $0.2$ & $0.5$ & $0.9$ \\ \hline
1 & 0.828 & 0.826 & 0.789 & 0.553 & 0.325 \\
2 & 0.618 & 0.615 & 0.565 & 0.439 & 0.434 \\
3 & 0.481 & 0.477 & 0.437 & 0.416 & 0.416 \\
4 & 0.390 & 0.387 & 0.367 & 0.367 & 0.367 \\
5 & 0.328 & 0.325 & 0.320 & 0.320 & 0.320 \\
6 & 0.282 & 0.281 & 0.280 & 0.280 & 0.280
\end{tabular}
\caption{The minimum radius of curvature $\ell/r$, defined in \eqref{ell}, as a function of $k$ and of the aspect ratio $\varepsilon=w/2r$ (maximum evaluated on the quadrature points of the mesh of Sect.~V).}
\label{tab:ell}
\end{table}

\subsubsection{One constant approximation}
Within the one constant approximation ($k_1= k_2=k_3= K$), we obtain:
\be
\begin{split}
2 \Vp = {K}\bigg[&(\alpha_{,u})^2 + \left(\frac{2 \alpha_{,v} }{f_v} + \frac{\kappa_v}{f_v^2}\right)^2 - \frac{c_v^2}{2} \cos 2 \alpha\\
&- \frac{2 c_v}{f_v^2} k r \sin 2 \alpha + \frac{c_v^2}{2} + 4 \frac{k^2 r^2}{f_v^4} \bigg],
\end{split}
\label{2w}
\ee
whence
\be
\begin{split}
2\Vp = {K}\bigg[&|\grads \alpha|^2  + 2 \frac{\kappa_v}{f_v^2} \ev_v \cdot \grads \alpha - \frac{c_v^2}{2} \cos 2 \alpha\\
&- \frac{2 c_v}{f_v^2} k r \sin 2 \alpha+ \left(\frac{\kappa_v}{f_v^2}\right)^2 + \frac{c_v^2}{2} + 4 \frac{k^2 r^2}{f_v^4}\bigg].
\end{split}
\ee
In the one constant approximation $2\Vp=K|\grads\nv|^2=K\left(|\mathcal D\nv|^2+|\Lv\nv|^2\right)$, where $\mathcal D\nv$ is the covariant derivative. The first contribution, $|\mathcal D\nv|^2=\kappa_n^2+\kappa_t^2=(\alpha_{,u})^2+\left(2\alpha_{,v}/f_v+\kappa_v/f_v^2\right)^2$, depends only on the intrinsic geometry, whereas the second one, $|\Lv\nv|^2=\tau_n^2+c_n^2=\lambda_+^2\cos^2\theta+\lambda_-^2\sin^2\theta$, with $\theta$ the angle between $\nv$ and the principal direction associated with $\lambda_+$, penalises the alignment of the director with the principal direction of larger curvature in modulus. The extrinsic contribution is a source term in the Euler--Lagrange equation below, since it depends on $\alpha$ and, because $K_G\neq 0$ for $k\neq 0$, it cannot vanish identically.

\section{Boundary value problem}
Within the one constant approximation the equilibrium configurations are the critical points of the energy functional
\be
\mathcal F[\alpha]=\frac K2\int_{-w/2}^{w/2}\int_0^{2\pi}\Big[\,\cdots\Big]\,\frac{f_v}{2}\,\dd v\,\dd u ,
\label{F}
\ee
where the square bracket is the one in \eqref{2w}, subject to the boundary conditions and to the continuity conditions at the identification edge specified below.

The Euler-Lagrange equation is
\be
\Delta_s \alpha +  \underbrace{\ev_v \cdot \grads \left(\frac{\kappa_v}{f_v^2}\right)}_{=\frac{2}{f_v}\,\partial_v\!\left(\frac{\kappa_v}{f_v^2}\right)}  - \frac{c_v^2}{2} \sin 2 \alpha  + \frac{2 c_v}{f_v^2} k r \cos 2 \alpha=0
\label{eq:el}
\ee
with
\be
\Delta_s \alpha := \alpha_{,uu} + \frac{\kappa_v}{f_v^2} \alpha_{,u}  - 4 \frac{f_{v,v}}{f_v^3}  \alpha_{,v} +  \frac{4}{f_v^2} \alpha_{,vv} ,
\ee
the Laplace--Beltrami operator of the metric \eqref{metric}. For $k=0$ equation \eqref{eq:el} reduces to $\alpha_{,uu}+\alpha_{,vv}/r^2=\sin 2\alpha/(2r^2)$, i.e.\ to the equation for a nematic on a cylinder of radius $r$, whose minimisers are the uniform states with $\nv$ along the axis.

\subsection{Boundary conditions}
This equation is supported by free boundary conditions (note that $\ev_u$ is the in-plane normal to the boundary), which we impose by requiring the vanishing of the tangential component of the gradient of the director along the normal:
\be
\tv\cdot(\grads \nv)\ev_u |_{u=\pm\frac{w}{2}} = 0
\ee
which leads to
\be
\alpha_{,u}\left(\pm\frac{w}{2},v \right) = 0.
\label{bc:free}
\ee
We remark that the vector $(\grads\nv)\ev_u=\alpha_{,u}\,\tv+\sin\alpha\,L_{uv}\,\bnu$ has also a normal component, $-\left(2kr/f_v^2\right)\sin\alpha\,\bnu$, which is fixed by the geometry and cannot be annihilated by the director field unless $\sin\alpha=0$. Consequently the free boundary condition cannot be formulated as the vanishing of the whole vector $(\grads\nv)\ev_u$. Condition \eqref{bc:free} is the natural boundary condition of \eqref{F}, since $\alpha_{,u}$ enters the energy density only through $\alpha_{,u}^2$, and it corresponds to the vanishing of the generalised contact force.

\paragraph{Unequal Frank constants.} For unequal Frank constants the natural boundary condition is obtained by requiring that the boundary term arising from the first variation of $\mathcal F$ vanish. We detail the derivation. Let $B:=2\alpha_{,v}/f_v+\kappa_v/f_v^2$, so that
\be
\kappa_t=-\sin\alpha\,\alpha_{,u}+\cos\alpha\,B,\qquad \kappa_n=\cos\alpha\,\alpha_{,u}+\sin\alpha\,B .
\ee
Only $\kappa_t$ and $\kappa_n$ depend on $\alpha_{,u}$, with $\partial\kappa_t/\partial\alpha_{,u}=-\sin\alpha$ and $\partial\kappa_n/\partial\alpha_{,u}=\cos\alpha$, whereas $\tau_n$ and $c_n$ depend on $\alpha$ alone. Therefore
\be
\frac{\partial (2\Vp)}{\partial\alpha_{,u}}=-2k_1\kappa_t\sin\alpha+2k_3\kappa_n\cos\alpha ,
\ee
and the first variation of \eqref{F} contains the boundary term $\int_0^{2\pi}\frac{\partial \Vp}{\partial\alpha_{,u}}\,\delta\alpha\,\frac{f_v}{2}\,\dd v\,\big|_{u=-w/2}^{u=w/2}$, in which $f_v>0$ and $\delta\alpha$ is arbitrary. Hence $\partial \Vp/\partial\alpha_{,u}=0$ at $u=\pm w/2$; substituting the expressions of $\kappa_t$ and $\kappa_n$ and collecting the terms in $\alpha_{,u}$ and in $B$ we obtain
\be
\begin{split}
&-k_1\sin\alpha\left(-\sin\alpha\,\alpha_{,u}+\cos\alpha\,B\right)\\
&\quad+k_3\cos\alpha\left(\cos\alpha\,\alpha_{,u}+\sin\alpha\,B\right)=0 ,
\end{split}
\ee
that is
\be
\begin{split}
&\left(k_1\sin^2\alpha+k_3\cos^2\alpha\right)\alpha_{,u}\\
&+(k_3-k_1)\sin\alpha\cos\alpha\left(\frac{2\alpha_{,v}}{f_v}+\frac{\kappa_v}{f_v^2}\right)\Big|_{u=\pm\frac w2}=0 .
\end{split}
\label{bc:aniso}
\ee
The coefficient of $\alpha_{,u}$ is strictly positive, so that \eqref{bc:aniso} can be solved for $\alpha_{,u}$,
\be
\alpha_{,u}=-\frac{(k_3-k_1)\sin\alpha\cos\alpha}{k_1\sin^2\alpha+k_3\cos^2\alpha}\,B\Big|_{u=\pm\frac w2},
\ee
and it reduces to \eqref{bc:free} for $k_1=k_3$. When $k_1\neq k_3$ the free boundary condition is of oblique type: the normal derivative $\alpha_{,u}$ at the edge is driven by the quantity $B$, which contains the tangential derivative $\alpha_{,v}$ and the geodesic curvature $\kappa_v/f_v^2$ of the edge, and it vanishes only where the director is parallel or orthogonal to the edge ($\sin\alpha\cos\alpha=0$) or where $B=0$. The anisotropy thus couples $\alpha_{,u}$ and $\alpha_{,v}$ through a factor proportional to $k_3-k_1$ and modulated by $\sin 2\alpha$: the director at the edge tilts so as to relax the more expensive of the two modes, splay for $k_1>k_3$ and bend for $k_1<k_3$.

\subsection{Conditions at the identification edge}
We recall that throughout the paper head-to-tail symmetry is implemented through the second-rank tensor $\Nv=\nv\otimes\nv$. In the plane tangent to the strip, with projector $\Pv$, this is the uniaxial $\Qv$-tensor $\Qv=S\,(\Nv-\tfrac12\Pv)$ of constant scalar order parameter $S>0$, i.e.\ the simplified form of the tensorial order parameter in which the degree of order is frozen and only the orientation is retained. Since $\Qv$ depends on $\nv$ only through $\Nv$, the statements ``the field is invariant under $\nv\leftrightarrow-\nv$'' and ``the tensor $\Nv$ is continuous'' are equivalent, and they are used interchangeably in what follows.
The formulation of continuity conditions at $v=0$ and $v=2\pi$ requires careful consideration, given the non-orientable nature of the surface. The conditions for joining the two ends of the domain at $v=0$ and $v=2\pi$ are expressed as follows:
\be
P (u, 0, \xi) = P((-1)^k u, 2 \pi, (-1)^k \xi)
\ee
and therefore
\be
P_\Sigma (u, 0) = P_\Sigma((-1)^k u, 2 \pi).
\ee
We note that the frame must be compared at the points that are identified, namely $(u,0)$ and $((-1)^ku,2\pi)$. Evaluating the expressions of Sect.~II.B at $v=0$ and $v=2\pi$ one finds $f_v\ev_v(u,0)=ku\,\ev_x+2r\,\ev_y$ and $f_v\ev_v(u,2\pi)=(-1)^kku\,\ev_x+2r\,\ev_y$, so that $\ev_v(u,2\pi)\neq\ev_v(u,0)$ for $k$ odd and $u\neq0$, while $\ev_v((-1)^ku,2\pi)=\ev_v(u,0)$ for every $k$. Likewise $\ev_u((-1)^ku,2\pi)=(-1)^k\ev_u(u,0)$ and hence $\bnu((-1)^ku,2\pi)=(-1)^k\bnu(u,0)$.

\subsection{$k$ is even}
When $k$ is even, the local frame $\{\ev_u, \ev_v, \bnu\}$ is continuous at the identification edge. Thus, taking into account the head-to-tail symmetry, one must require the continuity of the tensor field $\Nv= \nv \otimes \nv$, which translates into the condition
\be
\alpha(u,0) = \alpha(u, 2\pi) + p \pi, \qquad p \in \mathbb{Z}.
\label{bc11}
\ee
Furthermore, it is required that $\grads \Nv$ be continuous throughout the domain. This introduces an additional condition, namely that $(\grads \Nv) \ev_v$ be continuous at the identification edge. Expanding this condition, we obtain:
\be
(\grads \Nv) \ev_v = (\grads \nv) \ev_v \ot \nv + \nv \ot (\grads \nv) \ev_v
\ee
which, once \eqref{bc11} is taken into account, translates into a condition on the derivatives of $\alpha$:
\be
\alpha_{,v}(u,0) = \alpha_{,v}(u, 2\pi).
\label{bc12}
\ee
Differentiating \eqref{bc11} with respect to $u$ gives in addition $\alpha_{,u}(u,0)=\alpha_{,u}(u,2\pi)$, which expresses the continuity of $(\grads\Nv)\ev_u$.

\subsection{$k$ is odd}
Given that $\nv$ is defined via $\alpha$ within the local frame $\{\ev_u, \ev_v, \bnu\}$ which is not continuous across the identification edge, the continuity condition is rather less straightforward than in the previous case. In particular, we have
\[
\{\ev_u, \ev_v\}(u,0) = \{-\ev_u, \ev_v\}(-u,2\pi)
\]
and this leads to the conclusion that $\Nv$ is continuous if and only if
\be
\alpha(u,0) = - \alpha(-u, 2\pi) + p \pi, \qquad p \in \mathbb{Z}.
\label{bc21}
\ee

On the other hand, once equation \eqref{bc21} is taken into account, the condition of continuity of $\grads \Nv$ at the identification edge is obtained by requiring
\be
\alpha_{,v}(u,0) = -\alpha_{,v}(-u, 2\pi).
\label{bc22}
\ee
Differentiating \eqref{bc21} with respect to $u$ we obtain, moreover, $\alpha_{,u}(u,0)=\alpha_{,u}(-u,2\pi)$.
Finally, we can summarise the identification edge continuity conditions as follows:
\begin{subequations}
\be
\alpha(u,0) = (- 1)^k \alpha((-1)^k u, 2\pi) + p \pi, \qquad p, k \in \mathbb{Z}
\ee
\be
\alpha_{,v}(u,0) =  (- 1)^k\alpha_{,v}( (- 1)^k u, 2\pi), \qquad k \in \mathbb{Z}
\ee
\be
\alpha_{,u}(u,0) =  \alpha_{,u}( (- 1)^k u, 2\pi), \qquad k \in \mathbb{Z}
\ee
\label{bc}
\end{subequations}

\begin{figure}[t]
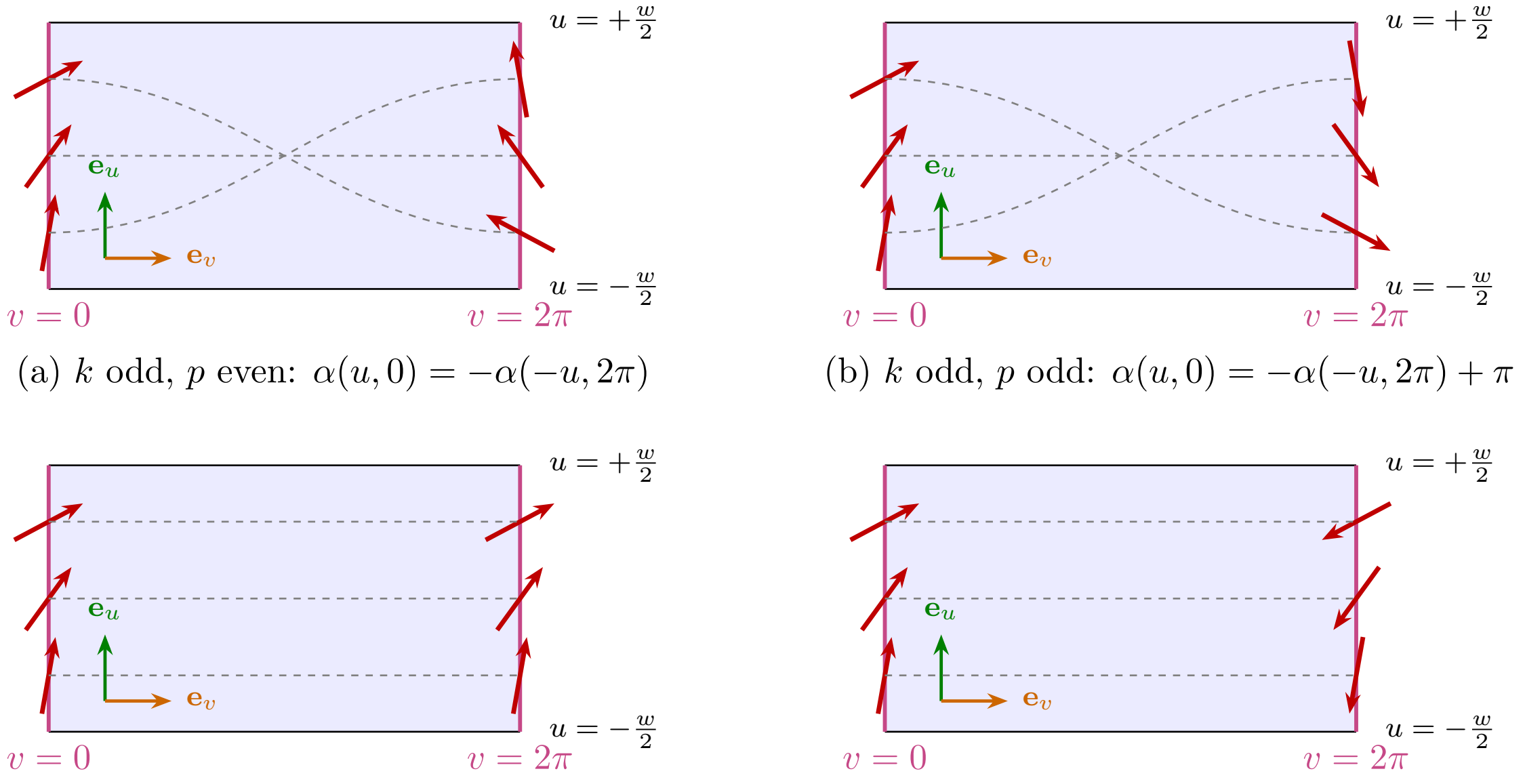

\centering
\begin{minipage}[t]{0.49\textwidth}\centering
\junction{1}{0}

\small (a) $k$ odd, $p$ even: $\alpha(u,0)=-\alpha(-u,2\pi)$
\end{minipage}\hfill
\begin{minipage}[t]{0.49\textwidth}\centering
\junction{1}{1}

\small (b) $k$ odd, $p$ odd: $\alpha(u,0)=-\alpha(-u,2\pi)+\pi$
\end{minipage}

\vspace{4mm}
\begin{minipage}[t]{0.49\textwidth}\centering
\junction{0}{0}

\small (c) $k$ even, $p$ even: $\alpha(u,0)=\alpha(u,2\pi)$
\end{minipage}\hfill
\begin{minipage}[t]{0.49\textwidth}\centering
\junction{0}{1}

\small (d) $k$ even, $p$ odd: $\alpha(u,0)=\alpha(u,2\pi)+\pi$
\end{minipage}
\caption{Continuity conditions at the identification edge. The strip is cut along $v=0\simeq2\pi$ and unrolled in the chart $(v,u)$, with $\mathbf e_v$ horizontal and $\mathbf e_u$ upward; the director is drawn as an arrow of angle $\alpha$ from $\mathbf e_u$, at three values of $u$ on the left edge ($\alpha=62^\circ,36^\circ,10^\circ$ from top to bottom) and at the glued points of the right edge. For $k$ odd the glued point of $(u,0)$ is $(-u,2\pi)$ and the frame of the right chart has $\mathbf e_u$ reversed, so that the arrow that continues the left one is its mirror image with respect to the horizontal axis; this is the case of the right edge in (b), where $\mathbf n$ is continuous as a vector, whereas in (a) it is replaced by $-\mathbf n$, which is the same nematic state. For $k$ even the glued point is $(u,2\pi)$, the frame is continuous, and the arrow is continuous in (c) and reversed in (d), in which $\alpha$ has changed by $\pi$ along the core line (a half turn of the director, $p=1$). In all cases $\mathbf N=\mathbf n\otimes\mathbf n$ is continuous, as required by \eqref{bc11} and \eqref{bc21}.}
\label{fig:junction}
\end{figure}

\subsection{Covariance of the problem under the identification}
The conditions \eqref{bc} are compatible with the equation \eqref{eq:el}. Indeed, a direct computation shows that, for $k$ odd,
\be
\begin{split}
&f_v(-u,v+2\pi)=f_v(u,v),\\
&\kappa_v(-u,v+2\pi)=-\kappa_v(u,v),\\
&c_v(-u,v+2\pi)=-c_v(u,v),
\end{split}
\ee
while $L_{uv}=-2kr/f_v^2$ is invariant, in agreement with the change of sign of $\ev_u$ and $\bnu$ at the identification edge; for $k$ even all these functions are $2\pi$-periodic in $v$. As a consequence, the energy density \eqref{2w} is invariant under the map $(u,v,\alpha)\mapsto(-u,v+2\pi,-\alpha+p\pi)$ for $k$ odd, and under $(u,v,\alpha)\mapsto(u,v+2\pi,\alpha+p\pi)$ for $k$ even, and the Euler--Lagrange equation \eqref{eq:el} is covariant under the same maps. Therefore, every solution of the problem on the strip can be lifted to a solution on the orientable double cover $v\in[0,4\pi)$ satisfying the corresponding symmetry relation.

\subsection{The role of the index $p$}
The integer $p$ has a different meaning in the two cases. Figure~\ref{fig:junction} illustrates the four cases.

For $k$ even, $p$ counts the number of half turns performed by the director with respect to the frame $\{\ev_u,\ev_v\}$ when the closed core line of the strip is covered once: since $p$ is an integer and varies continuously with $\alpha$, configurations with different values of $p$ cannot be connected by continuous deformations of $\alpha$ compatible with \eqref{bc}, and $p\in\mathbb Z$ is a topological index.

For $k$ odd, the transformation $\alpha\mapsto\alpha+\pi$ maps $\nv$ into $-\nv$ and therefore leaves the tensor $\Nv=\nv\otimes\nv$ unchanged, i.e.\ it describes the same nematic state; it maps $p$ into $p+2$. Therefore only the parity of $p$ is physically meaningful: the solutions split into two sectors, $p$ even and $p$ odd, which are not related by a continuous deformation compatible with \eqref{bc21}. In this case condition \eqref{bc21} constrains the sum $\alpha(u,0)+\alpha(-u,2\pi)$ and not the difference $\alpha(u,2\pi)-\alpha(u,0)$, so that the net rotation of the director along the core line is not quantised. In particular, the minimisation of the energy with respect to $p$ must be understood as a comparison between the sectors $p\in\{0,1\}$ for $k$ odd, whereas for $k$ even it is a comparison among all integers. This is the nematic counterpart, modulo~$\pi$, of the identification $\theta(x+L,y)=-\theta(x,w-y)$ used for XY models on a Klein bottle in Ref.~\cite{S26}, where it pins a homogeneous field along the twist axis (their angle is measured from the twist axis, ours from $\mathbf{e}_u$). Here a field constant in the frame $\{\mathbf{e}_u,\mathbf{e}_v\}$ is compatible with \eqref{bc21} for $\alpha_0=p\pi/2$, i.e.\ $\alpha_0=\pi/2$ for $p$ odd and $\alpha_0=0$ for $p$ even. For $p$ even the extrinsic coupling lifts it, since $\alpha_0=0$ does not satisfy \eqref{ode}; for $p$ odd $\alpha_0=\pi/2$ remains an exact solution, but it is a saddle of $\mathcal E$ (Sect.~IV.C).

\section{The narrow-strip limit}
\subsection{Reduced functional and reduced equation}
When $w\ll r$ the coefficients of \eqref{eq:el} may be evaluated at $u=0$, where $f_v=2r$, $\kappa_v/f_v^2=\sin(kv/2)/r$ and $c_v=\cos(kv/2)/r$, and the director angle is sought in the form $\alpha=\alpha(v)$, which is compatible with \eqref{bc:free}. The functional \eqref{F} then reduces to
\be
\begin{split}
&\mathcal F\simeq\frac{Kw}{2r}\,\mathcal E[\alpha],\\
&\mathcal E[\alpha]=\int_0^{2\pi}\Big[\Big(\alpha'+\sin\tfrac{kv}{2}\Big)^2+\cos^2\tfrac{kv}{2}\,\sin^2\alpha\\
&\qquad-\tfrac k2\cos\tfrac{kv}{2}\,\sin 2\alpha+\tfrac{k^2}{4}\Big]\dd v ,
\end{split}
\label{Ered}
\ee
where the prime denotes differentiation with respect to $v$ and the constant has been chosen so that $\mathcal E=0$ for the cylinder with $\nv$ along the axis. The radius $r$ enters only through the prefactor, and the sole control parameter of the reduced problem is the number of half-twists $k$. The corresponding Euler--Lagrange equation is
\be
\alpha''=\frac12\cos^2\frac{kv}{2}\,\sin 2\alpha-k\cos\frac{kv}{2}\,\cos^2\alpha ,
\label{ode}
\ee
and it is supplemented by the conditions \eqref{bc11}--\eqref{bc12} for $k$ even and \eqref{bc21}--\eqref{bc22} for $k$ odd, in which the dependence on $u$ drops out:
\be
\alpha(0)=(-1)^k\alpha(2\pi)+p\pi,\qquad \alpha'(0)=(-1)^k\alpha'(2\pi).
\ee
For $k=0$ equation \eqref{ode} is the pendulum equation $\alpha''=\frac12\sin 2\alpha$, whose solutions are expressed by Jacobi elliptic functions. This is the case of a nematic on a cylinder, studied in Ref.~\cite{NV12}, where only homogeneous solutions were considered; note that in that reference the angle is measured from the parallels, so that the director along the generatrices is $\alpha=\pi/2$ there, whereas here it is $\alpha=0$. In the present convention the uniform state $\alpha=0$ is the absolute minimiser of $\mathcal E$, with $\mathcal E=0$, since $\mathcal E=\int_0^{2\pi}(\alpha'^2+\sin^2\alpha)\,dv\ge0$ for $k=0$, and $\alpha=\pi/2$ (the director along the circles) is unstable. The sectors with $p\ge1$ are not homogeneous: the first integral of \eqref{ode} gives $\alpha'^2=\sin^2\alpha+C$, with $C>0$ fixed by the condition that $\alpha$ increases by $p\pi$ over a period, namely $p\,2K(m)/\sqrt{1+C}=2\pi$ with $m=1/(1+C)$, where $K$ is the complete elliptic integral of the first kind, and their energy is $\mathcal E=p\int_0^\pi(2\sin^2a+C)/\sqrt{\sin^2a+C}\,da$, which gives $4.03$, $9.23$ and $17.19$ for $p=1,2,3$, in agreement with Table~\ref{tab:E}.

Two elementary observations follow. First, $\alpha=\pi/2$, i.e.\ $\nv=\ev_v$ along the core line, solves \eqref{ode} for every $k$ and is compatible with the identification conditions for $k$ even with $p=0$ and for $k$ odd with $p$ odd, but it is a critical point of $\mathcal E$ that is not a minimiser, since its energy $2\pi+\pi k^2/2$ is higher than that of the minimisers reported below; the number of negative eigenvalues of the second variation of $\mathcal E$ at this point (Morse index) is $1$ for $k$ even and $2$ for $k$ odd, as we checked for $k\le4$. Second, $\alpha=0$ ($\nv=\ev_u$) is not a solution for $k\neq 0$, since the right-hand side of \eqref{ode} equals $-k\cos(kv/2)$ there. Consequently, for $k\neq0$ the geometric frustration is present already in the equilibrium problem.

\subsection{Accuracy of the reduction}
The reduced equation is exact only at leading order in $w/r$: since $f_v$, $\kappa_v$ and $c_v$ depend on $u$, a function $\alpha=\alpha(v)$ does not satisfy \eqref{eq:el} for every $u$. We evaluated the residual of the complete equation \eqref{eq:el} on the minimisers of \eqref{Ered}, in units in which $r=1$ and with $\alpha_{,u}=\alpha_{,uu}=0$. For $k=1$ and $p=1$ the maximum over $v$ of the residual is $1.2\times10^{-2}$, $2.3\times10^{-2}$ and $4.1\times10^{-2}$ for $|u|/r=0.05,\,0.1,\,0.2$, respectively, i.e.\ it grows linearly in $|u|/r$ and vanishes at $u=0$ (the same behaviour is found for $k=1$, $p=0$ and for $k=2$, $p=0$, with coefficients between $0.2$ and $0.65$). The magnitude of $\alpha''$ is $0.3$--$0.8$ in the same units. Therefore, solutions with a non-trivial dependence on $u$ are generated by the terms of order $|u|/r$, and their characterisation requires the solution of the complete problem, which is carried out in Sect.~V.

\subsection{Minimisers of the reduced functional}
\begin{figure}[t]
\centering
\includegraphics[width=\textwidth]{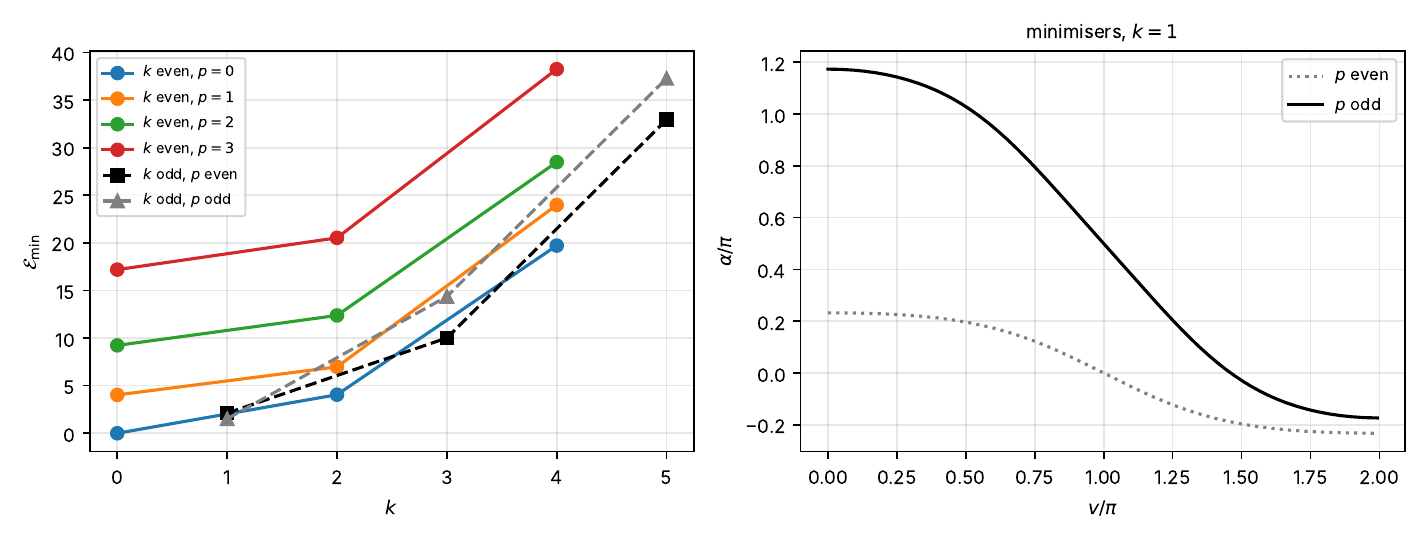}
\caption{Left: minimum of $\mathcal E$ versus $k$; for $k$ even the curves refer to $p=0,\dots,3$, for $k$ odd to the two sectors $p$ even and $p$ odd. Right: minimisers $\alpha(v)$ for $k=1$ in the sectors $p$ even and $p$ odd.}
\label{fig:red}
\end{figure}

We minimised $\mathcal E$ for $k=0,\dots,5$ and $p=0,\dots,3$ by a quasi-Newton method started from $120$ random initial conditions, and checked the results for $k=1,\dots,4$ by solving the boundary value problem \eqref{ode}, \eqref{bc} (App.~\ref{app:red}). The lowest energies found are collected in Table~\ref{tab:E} and Fig.~\ref{fig:red}; they do not exclude lower minima, and the boundary value problem shows that further local minima of higher energy exist in some sectors (for instance $E\simeq7.68$ for $k=1$, $p$ odd, and $E\simeq20.21$ for $k=3$, $p$ even).

\begin{table}[h]
\centering
\begin{tabular}{c|cccc}
\multicolumn{5}{c}{$k$ even}\\
$k$ & $p=0$ & $p=1$ & $p=2$ & $p=3$ \\ \hline
0 & 0.00 & 4.03 & 9.23 & 17.19 \\
2 & \textbf{4.05} & 6.97 & 12.38 & 20.53 \\
4 & \textbf{19.72} & 23.98 & 28.49 & 38.26
\end{tabular}

\vspace{2mm}
\begin{tabular}{c|cc}
\multicolumn{3}{c}{$k$ odd}\\
$k$ & $p$ even & $p$ odd \\ \hline
1 & 2.11 & \textbf{1.51} \\
3 & \textbf{10.02} & 14.37 \\
5 & \textbf{32.95} & 37.28
\end{tabular}
\caption{Minimum of the reduced functional $\mathcal E$ in \eqref{Ered} as a function of $k$ and $p$; the lowest value for each $k$ is in bold. For $k$ even (upper block) $p\in\mathbb Z$ is a winding number and the energy is tabulated for $p=0,\dots,3$; for $k=0$ the value $p=0$ is the cylinder with $\nv$ along the axis. For $k$ odd (lower block) the equivalence $p\sim p+2$ identifies the columns $p$ and $p+2$ (the minima for $p=0,2$ and for $p=1,3$ coincide to the numerical accuracy), so that only the parity of $p$ labels the sectors; since \eqref{bc21} constrains the sum $\alpha(u,0)+\alpha(-u,2\pi)$, the net rotation of the director along the core line is not quantised in this case.}
\label{tab:E}
\end{table}

The data of Table~\ref{tab:E} show that: (i) for $k$ even the minimum is attained at $p=0$ and the energy increases with $|p|$, namely a net winding of the director along the core line has an elastic cost; (ii) for $k$ odd the two sectors are not equivalent, and the sector selected by the energy depends on $k$: the odd sector is favoured for $k=1$, by a margin of $0.60$, whereas the even sector is favoured for $k=3$ and $k=5$, by margins of about $4.3$; (iii) the minimum energy grows with $k$, consistently with the cost of accommodating the twist of the strip.

\section{The complete two-dimensional problem}
The residual analysis of Sect.~IV.B shows that the reduced description is accurate only to leading order in $\varepsilon$. We therefore solve numerically the full variational problem for $\alpha(u,v)$, in the one constant approximation unless stated otherwise, and we determine the minimisers in each sector of the index $p$ as functions of $k$, $\varepsilon$ and of the ratios $k_1/k_3$, $k_2/k_3$. We emphasise that we compute minimisers (stable branches) and not the full set of critical points; the bifurcation structure of the unstable branches is not investigated.

\subsection{Method}
We set $r=1$ and we discretise the energy, normalised per unit width,
\be
\hat{\mathcal E}:=\frac1w\int_{-w/2}^{w/2}\!\!\int_0^{2\pi}\frac{2\Vp}{K}\,\frac{f_v}{2}\,\dd v\,\dd u ,
\label{Ehat}
\ee
where $\Vp$ is the Frank density of Sect.~II, so that $\hat{\mathcal E}\to\mathcal E$ as $\varepsilon\to0$ and $\mathcal F_{2D}/K=\varepsilon\,\hat{\mathcal E}$ for $r=1$. The functional is discretised with bilinear finite elements on a uniform mesh of $N_u\times N_v=16\times128$ cells in $(u,v)\in[-\varepsilon,\varepsilon]\times[0,2\pi)$, with the identification edge imposed exactly, and minimised by a quasi-Newton method, from the reduced minimiser continued in $\varepsilon$ and from random initial conditions of three types; the minima reported are the lowest found, not proved to be global. The discretisation, the convergence tests and the independent checks are described in App.~\ref{app:2D}.

\subsection{Energy as a function of the aspect ratio}
\begin{figure}[t]
\centering
\includegraphics[width=\textwidth]{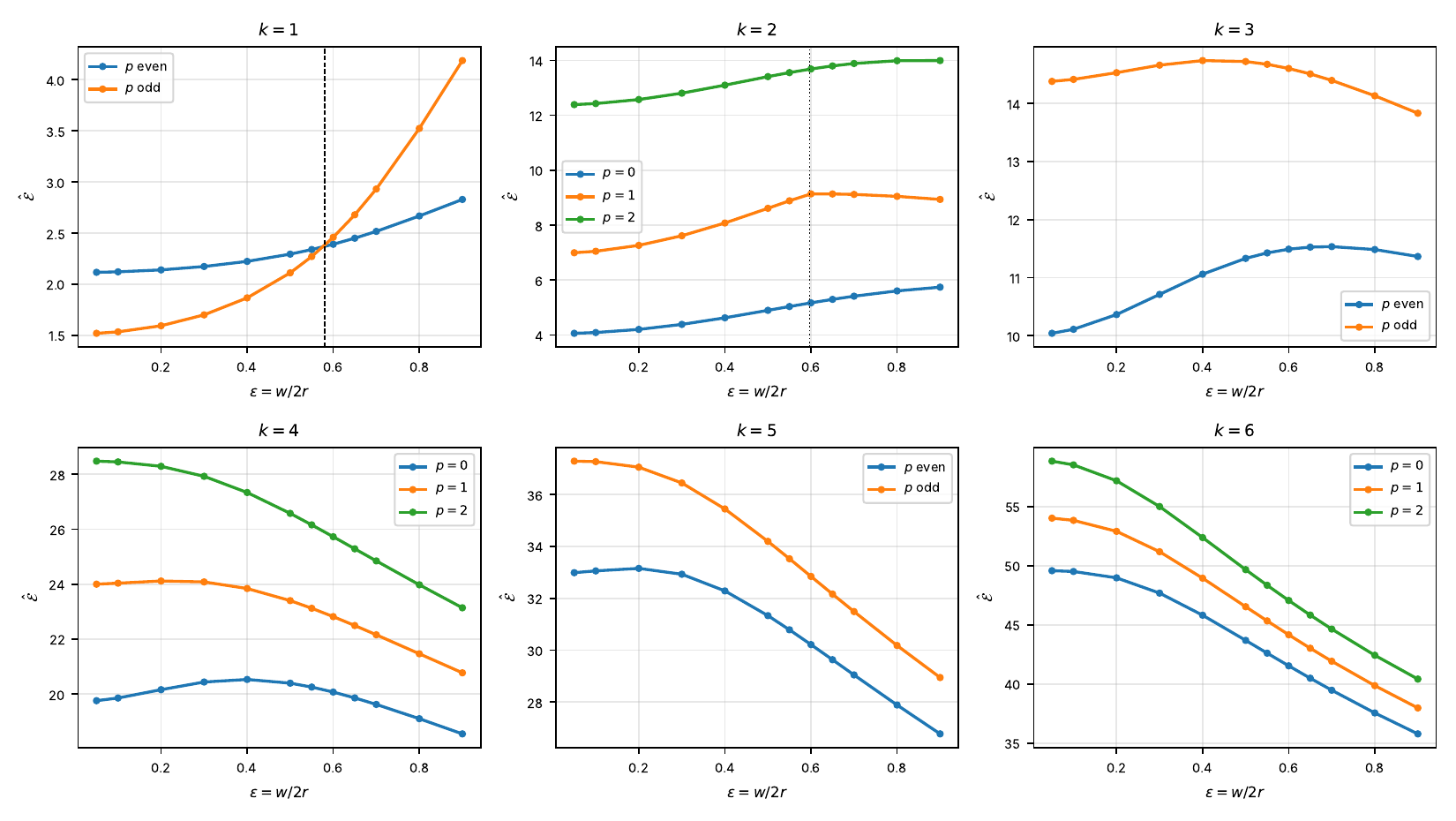}
\caption{Minimum of $\hat{\mathcal E}$ in each sector as a function of $\varepsilon=w/2r$, for $k=1,\dots,6$ (isotropic case). For $k$ odd the two sectors are $p$ even and $p$ odd; for $k$ even the curves refer to $p=0,1,2$ (the sectors $\pm p$ are degenerate). The dashed line for $k=1$ marks the crossing $\varepsilon_c\simeq0.58$; for $k=2$, $p=1$ the curve is the lower envelope of the two branches of Table~\ref{tab:k2p1}, which cross at $\varepsilon_*\simeq0.60$ (dotted line).}
\label{fig:eps}
\end{figure}
Figure~\ref{fig:eps} shows that, for every $k$ and $p$, $\hat{\mathcal E}$ varies smoothly with $\varepsilon$, with the exception of the sector $k=2$, $p=1$ discussed below, and tends to the reduced value for $\varepsilon\to0$. The only crossing between sectors found for $k\le6$ in the explored range $0.05\le\varepsilon\le0.9$ occurs for $k=1$: the odd sector, which is favoured in the narrow-strip limit, loses to the even sector for $\varepsilon>\varepsilon_c\simeq0.58$ (Table~\ref{tab:k1}). In the same range the sector selected for $k=2,\dots,6$ is the one of the reduced problem: $p=0$ for $k$ even and $p$ even for $k$ odd, $k\ge3$.
The exception is the sector $k=2$, $p=1$, which for $0.3\le\varepsilon\le0.9$ has (at least) two distinct minima, both with positive definite Hessian, that we denote by A and B (Table~\ref{tab:k2p1}). A is the branch obtained by continuation from the reduced solution; it is the lowest for $\varepsilon<\varepsilon_*$, whereas B is the lowest for $\varepsilon>\varepsilon_*$, where $\varepsilon_*\simeq0.60$ (linear interpolation of $\hat{\mathcal E}_{\rm A}-\hat{\mathcal E}_{\rm B}$ between $\varepsilon=0.55$ and $0.60$ gives $0.597$). The two branches do not merge, so that there is no bifurcation: the minimum energy of the sector, $\min\{\hat{\mathcal E}_{\rm A},\hat{\mathcal E}_{\rm B}\}$, is continuous and has a change of slope at $\varepsilon_*$. The branch B has the smaller distortion, $\max|\grads\nv|\simeq2.4/r$ against $3.3$--$3.4/r$, and the larger variation of $\alpha$ across the width ($8.5^\circ$ against $7.7^\circ$ at $\varepsilon=0.9$). The two branches are found with comparable frequency from random starts near $\varepsilon_*$ (for instance $14$ and $46$ out of $60$ at $\varepsilon=0.6$, where the lowest minimum is the less frequent one), so that a handful of starts may miss the lowest. No other sector with $k\le6$ shows such an exchange on the grid of twelve values of $\varepsilon$ between $0.05$ and $0.9$ that we used (the second minimum lies at least $0.9$ above the lowest). The change of branch does not alter the sector selected for $k=2$, since the gap to the sector $p=0$ remains larger than $3.2$.
\begin{table}[h]
\centering
\begin{tabular}{c|cccccc}
$\varepsilon$ & 0.3 & 0.5 & 0.6 & 0.7 & 0.8 & 0.9 \\ \hline
A & \textbf{7.613} & \textbf{8.613} & 9.155 & 9.642 & 10.030 & 10.305 \\
B & 8.941 & 9.103 & \textbf{9.137} & \textbf{9.119} & \textbf{9.048} & \textbf{8.936}
\end{tabular}
\caption{The two minima A and B of $\hat{\mathcal E}$ for $k=2$, $p=1$ as functions of $\varepsilon$ (isotropic case); the lowest value is in bold. The branches exchange their order at $\varepsilon_*\simeq0.60$.}
\label{tab:k2p1}
\end{table}
\begin{table}[t]
\centering
\small
\begin{tabular}{c|cccccccccc}
$\varepsilon$ & 0.05 & 0.1 & 0.2 & 0.3 & 0.4 & 0.5 & 0.6 & 0.7 & 0.8 & 0.9 \\ \hline
$p$ even & 2.116 & 2.120 & 2.139 & 2.172 & 2.223 & 2.294 & 2.390 & 2.516 & 2.667 & 2.828 \\
$p$ odd & 1.519 & 1.533 & 1.593 & 1.699 & 1.866 & 2.110 & 2.458 & 2.931 & 3.523 & 4.188
\end{tabular}
\caption{Minimum of $\hat{\mathcal E}$ for $k=1$ in the two sectors, as a function of $\varepsilon$; the sectors cross at $\varepsilon_c\simeq0.58$.}
\label{tab:k1}
\end{table}
The energy of the odd sector grows much faster with $\varepsilon$ than that of the even sector (by a factor $2.8$ and $1.3$, respectively, between $\varepsilon=0.05$ and $0.9$); we have not derived an analytical explanation of this difference, which we regard as a numerical finding.

\subsection{Phase diagram in the plane $(k,\varepsilon)$}
\begin{figure}[t]
\centering
\includegraphics[width=0.85\textwidth]{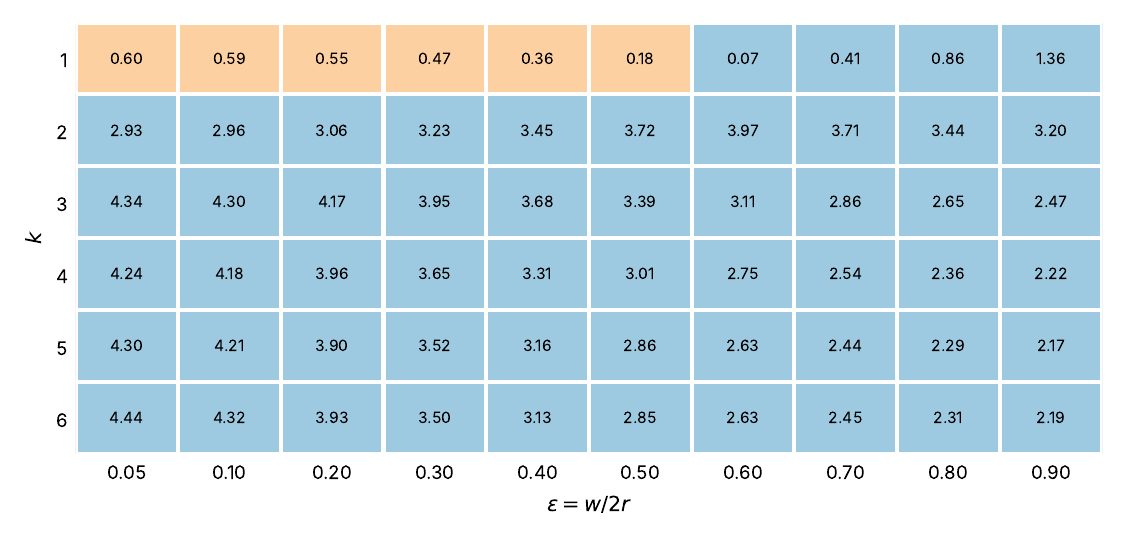}
\caption{Phase diagram in the plane $(k,\varepsilon)$ for the isotropic energy. The colour marks the sector of minimal energy (blue: $p=0$, equivalently $p$ even, as in Fig.~\ref{fig:eps}; orange: $p$ odd, possible for $k$ odd only); the number in each cell is the energy gap $\Delta\hat{\mathcal E}$ between the lowest and the second lowest sector.}
\label{fig:phase}
\end{figure}
The map of the sector of minimal energy in the plane $(k,\varepsilon)$ is reported in Fig.~\ref{fig:phase}, together with the gap to the second lowest sector. Apart from the cell $k=1$, $\varepsilon\gtrsim0.58$, the ordering of the sectors of the reduced problem is preserved up to $\varepsilon=0.9$. The gap behaves differently in the two parities: for $k=2$ it increases from $2.9$ at $\varepsilon=0.05$ to a maximum of about $4.0$ at $\varepsilon\simeq0.6$ and then decreases to $3.2$ at $\varepsilon=0.9$, because of the change of branch in the sector $p=1$ (Table~\ref{tab:k2p1}), whereas for $k=3,\dots,6$ it decreases from about $4.3$ to $2.2$ ($k=4,5,6$) or $2.5$ ($k=3$); for $k=1$ it is non-monotonic, with a minimum $0.07$ close to the crossing. The sector selection is therefore robust for $k\ge2$ in the isotropic case and it is delicate for $k=1$, the physically most relevant case.

\subsection{Dependence on the transverse coordinate}
\begin{figure}[t]
\centering
\includegraphics[width=\textwidth]{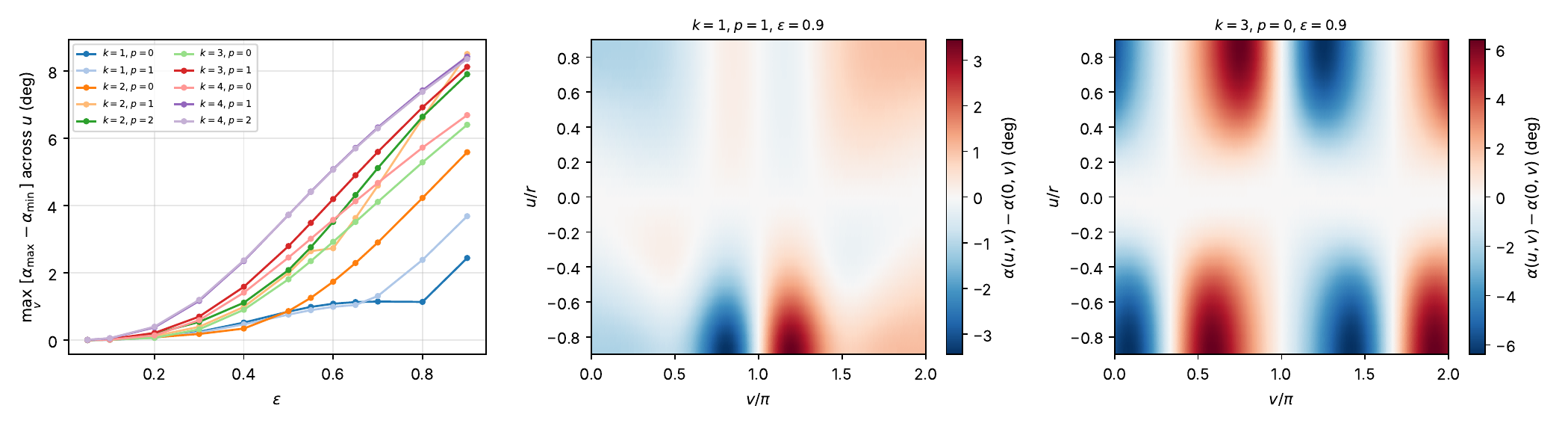}
\caption{Left: maximum over $v$ of the variation of $\alpha$ across the strip, $\alpha_{\max}-\alpha_{\min}$ at fixed $v$, as a function of $\varepsilon$ for the computed minimisers with $k=1,\dots,4$ (the values for $k=5,6$ are given in the text). Centre and right: the deviation $\alpha(u,v)-\alpha(0,v)$ for $k=1$, $p$ odd, and for $k=3$, $p$ even, at $\varepsilon=0.9$. The curves are those of the lowest minima on the mesh $16\times128$ (for $k=2$, $p=1$ the lower envelope of the two branches of Table~\ref{tab:k2p1}); the maps are computed on the mesh $32\times256$ and plotted with the bilinear interpolation of the nodal values.}
\label{fig:u}
\end{figure}
The question raised in Sect.~IV.B, namely whether solutions with a non-trivial dependence on $u$ exist, has a precise answer in the isotropic case: the minimisers do depend on $u$, but the dependence is weak and perturbative. The variation of $\alpha$ across the width never exceeds $0.147$ rad ($8.4^\circ$; $8.5^\circ$ with the spectral code for $k=4$, $p=1$, $\varepsilon=0.9$, and $8.5^\circ$ with the finite elements for $k=2$, $p=1$, $\varepsilon=0.9$ on the branch B of Sect.~V.B) for any computed isotropic minimiser with $k\le6$ and $\varepsilon\le0.9$, and it is at most $2.4^\circ$ ($k=1$), $5.6^\circ$ ($k=2$), $6.4^\circ$ ($k=3$) and $6.8^\circ$ ($k=4,5,6$) in the sector $p=0$ (Fig.~\ref{fig:u}). No secondary branch with a large-amplitude modulation in $u$ was found among the minimisers, starting either from the reduced solution or from random fields. These observations suggest that the large variation of $\hat{\mathcal E}$ with $\varepsilon$ in Fig.~\ref{fig:eps} is due predominantly to the dependence on $u$ of the geometric coefficients, and not to a genuine two-dimensional texture; we have not, however, tested this by evaluating the full energy on the $u$-independent ansatz.

\subsection{Elastic anisotropy}
\begin{figure}[t]
\centering
\includegraphics[width=\textwidth]{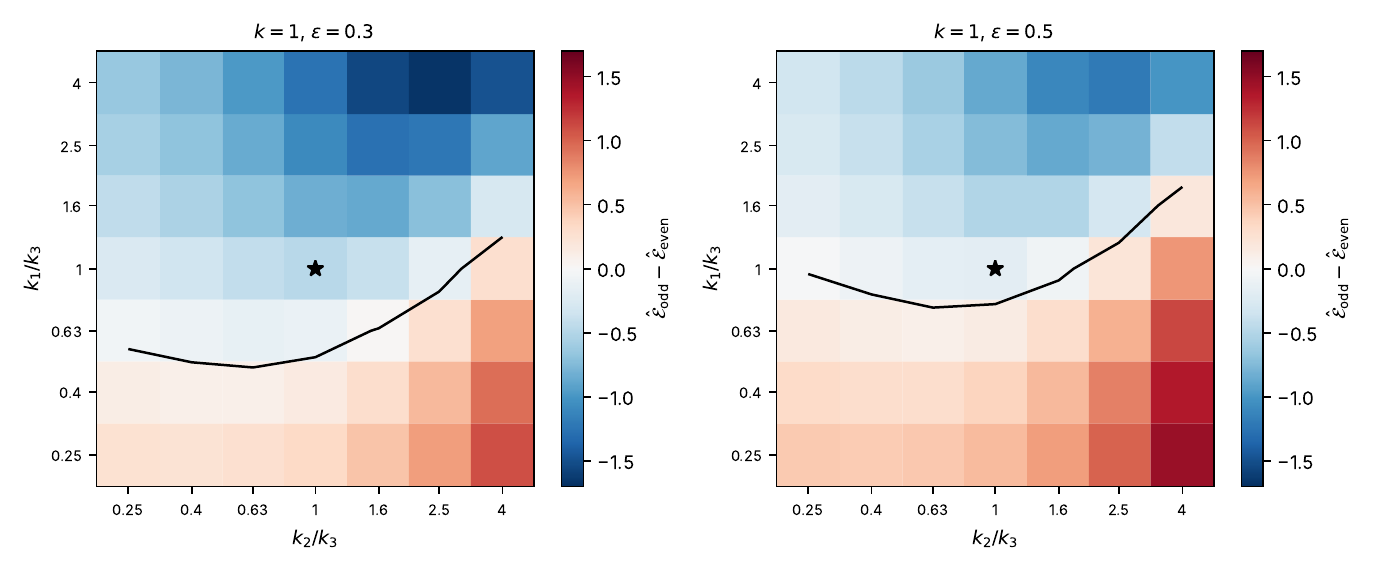}
\caption{Energy difference $\hat{\mathcal E}_{\rm odd}-\hat{\mathcal E}_{\rm even}$ between the two sectors for $k=1$ in the plane $(k_2/k_3,k_1/k_3)$, for $\varepsilon=0.3$ (left) and $\varepsilon=0.5$ (right), on a $7\times7$ logarithmic grid between $1/4$ and $4$; the black line is the contour of zero difference (phase boundary), and the star marks the isotropic point; each cell corresponds to a distinct minimisation, so that the colour is piecewise constant by construction.}
\label{fig:aniso}
\end{figure}
For $k=1$ we computed the minima of the two sectors for $k_1/k_3$ and $k_2/k_3$ varying on a $7\times7$ logarithmic grid in $[1/4,4]$ at $\varepsilon=0.3$ and $\varepsilon=0.5$, with the boundary condition \eqref{bc:aniso} (Fig.~\ref{fig:aniso}). At $\varepsilon=0.3$ the odd sector is favoured in $63\%$ of the grid points and at $\varepsilon=0.5$ in $51\%$, in agreement with the narrowing of the isotropic gap at the isotropic point ($-0.47$ at $\varepsilon=0.3$, $-0.18$ at $\varepsilon=0.5$). The phase boundary is controlled mainly by $k_1/k_3$: a stiff splay constant favours the odd sector, since for $k_1/k_3=4$ the difference is negative for every $k_2/k_3\le4$, whereas for $k_1/k_3=1/4$ the even sector is favoured at every value of $k_2/k_3$ explored. A large twist constant favours the even sector when $k_1/k_3\lesssim1$ (for instance, for $k_2/k_3=4$ at $k_1/k_3=1$ and $\varepsilon=0.3$ the difference is $+0.29$). Increasing $\varepsilon$ displaces the boundary towards larger $k_1/k_3$, consistently with the isotropic crossing at $\varepsilon_c\simeq0.58$. In the whole grid the largest distortion $\max|\grads\nv|$ is $1.8/r$ at $\varepsilon=0.3$ and $2.9/r$ at $\varepsilon=0.5$, and the variation of $\alpha$ across the width stays below $15^\circ$ and $21^\circ$, respectively. Values of the ratios outside $[1/4,4]$, which are relevant for lyotropic and polymeric nematics, have not been explored.

\subsection{Mirror symmetry and comparison with the ferromagnetic case}
The energy density is invariant under the map $(k,u,\alpha)\mapsto(-k,-u,-\alpha)$, together with $p\mapsto-p$ in the identification conditions, namely $\alpha_{-k,-p}(u,v)=-\alpha_{k,p}(-u,v)$; the reflection of the transverse coordinate is essential, since for $u\neq0$ the replacement $k\to-k$ alone changes $f_v$ and $c_v$ through $\sin(kv/2)\to-\sin(kv/2)$. Indeed, one verifies directly that $f_v(-k,-u,v)=f_v(k,u,v)$ and $c_v(-k,-u,v)=c_v(k,u,v)$, whereas $\kappa_v(-k,-u,v)=-\kappa_v(k,u,v)$ and $L_{uv}\to-L_{uv}$. Setting $\tilde\alpha(u,v):=-\alpha(-u,v)$, one obtains $\tilde\alpha_{,v}=-\alpha_{,v}(-u,v)$, $\tilde\alpha_{,u}=\alpha_{,u}(-u,v)$, and therefore $\kappa_t\to-\kappa_t$, $\kappa_n\to\kappa_n$, $\tau_n\to-\tau_n$, $c_n\to c_n$, so that each of the four squares in the Frank density, and hence the energy for arbitrary $k_1,k_2,k_3$, is invariant; the domain $u\in[-w/2,w/2]$ and the free boundary conditions \eqref{bc:free}, \eqref{bc:aniso} are mapped into themselves, and \eqref{bc} is preserved with $p\to-p$ (for $k$ odd only the parity of $p$ matters). At $u=0$ the map reduces to $(k,\alpha)\mapsto(-k,-\alpha)$, which is the symmetry of the reduced functional \eqref{Ered}. We verified numerically at $\varepsilon=0.3$ that the minimum energies coincide, to $10^{-12}$ in absolute value (the relative difference is below $10^{-15}$ in the independent implementation of App.~\ref{app:2D}, where the absolute differences are between $10^{-15}$ and $9\times10^{-15}$), for $k=\pm1$ and $k=\pm2$ in the sectors $p=0,1$, and that for $k=2$ the sectors $p=1,2$ and $p=-1,-2$ are degenerate. With unequal constants ($k_1:k_2:k_3=1.3:0.7:1$) the spectral code gives the same coincidence to $10^{-13}$ for $k=\pm1,\pm2$ and $p=0,\pm1$, and also the equivalence $p\sim p+2$ for $k=1,3$. Hence the energy does not distinguish between the two enantiomorphic strips, and the solutions do not select a handedness of the director relative to that of the strip. This is a consequence of the achirality of the Frank energy, whose extrinsic part depends on $\Lv$ only through $|\Lv\nv|^2$, which is even under $\bnu\to-\bnu$ (for unequal constants through the squares of the two components of $\Lv\nv$, along $\nv$ and along $\bnu\times\nv$, weighted by $k_3$ and $k_2$, both even under $\bnu\to-\bnu$). It contrasts with the ferromagnetic Möbius ring of \cite{PK15}, where the exchange energy produces curvature- and torsion-induced terms linear in $\Lv$ and, hence, a coupling between the chirality of the ring and that of the magnetic texture; for a director constrained to be tangent to the strip no such coupling arises in the Frank energy, as shown by the invariance above. In a nematic the analogous coupling would be produced by a chiral cholesteric term in the Frank energy, which we have not included.

\subsection{Validity of the smooth description}
The Frank energy used above describes smooth line fields and is justified when the scale of variation of $\nv$ is large compared with the nematic coherence length $\xi$, i.e.\ $\xi\max|\grads\nv|\ll1$. In all the computed minimisers the largest distortion is $\max|\grads\nv|\le4.3/r$ (attained for $k=6$), so that the condition holds for $\xi\ll r/4.3$; this is a condition on the thickness scale, not on the topology, and it does not preclude the nucleation of defects.

The hypotheses of Sect.~II do not allow for defects. The director is tangent to the strip, $\nv\cdot\bnu=0$, which excludes the escape of the director along the thickness and the removal of the cores, and the topology does not require them: $\chi(\Sigma)=0$, a smooth line field exists in every sector, and the minimisers computed here are smooth. As a check of the geometry, the Gauss--Bonnet identity $\int_\Sigma K_G\,\dd A+\oint_{\partial\Sigma}k_g\,\dd s=2\pi\chi(\Sigma)=0$, with $\oint_{\partial\Sigma}k_g\,\dd s=\int_0^{2\pi}\big[\partial_uh\big]_{u=-w/2}^{u=w/2}\dd v$ and $h=f_v/2$ in the metric \eqref{metric}, was verified numerically; for instance $\oint_{\partial\Sigma}k_g\,\dd s/2\pi=2.025=-Q_K$ for $k=3$, $\varepsilon=0.5$, with $Q_K=(2\pi)^{-1}\int_\Sigma K_G\,\dd A$.

The nucleation of disclination pairs, which for a free edge could also occur through the edge, is not considered, and its study requires a tensorial order parameter $\Qv$ with variable scalar order. Two consequences are relevant for the results above. First, since the edges are free, the index $p$, a property of smooth fields, is not conserved if a half-defect leaves the strip, so that the sector structure of Sect.~III holds for smooth textures and the sector gaps of Fig.~\ref{fig:phase} (at least $2.2$ for $k\ge2$) are not energy barriers. Second, the elastic energy of the smooth minimisers grows with $k$ (Table~\ref{tab:E}) and with $\varepsilon$, whereas the energy of a pair $\pm\tfrac12$ at distance $d\gg\xi$ grows only as $\tfrac\pi2K\ln(d/\xi)$ in the one constant approximation, so that for large $k$ and large $\varepsilon$ a texture with defects may have a lower energy than the smooth one: the minimisers with $k\ge4$ are to be read as the lowest smooth states and not as the ground states of the nematic film. We do not attempt to quantify this regime.

\section{Concluding remarks}
To the best of our knowledge, the equilibrium problem of a nematic confined in a tubular neighbourhood of a M\"obius strip, that is, of a nematic reduced to the mid-surface of the strip, with a director tangent to it, with the full extrinsic Frank energy and with identification conditions that account for the non-orientability of the surface and for the head-to-tail symmetry of the director, had not been addressed before; the results above are, in this sense, original. The parity of the number $k$ of half-twists determines the nature of the topological index $p$. For $k$ even, $p$ is a genuine winding number, and configurations with different $p$ cannot be deformed into one another compatibly with the identification conditions; for $k$ odd, $\alpha\mapsto\alpha+\pi$ describes the same nematic state, only the parity of $p$ is physical, and the net rotation of the director along the core line is not quantised. This distinction, a consequence of the non-orientability of the strip combined with the symmetry $\nv\to-\nv$, is the main structural result of the paper.

In the narrow-strip limit the problem reduces to an ordinary boundary value problem. For $k=0$ it is the pendulum equation of the cylindrical strip \cite{NV12}: the uniform state $p=0$ has zero energy, and the solutions with a net winding $p=1,2,3$ cost $4.03$, $9.23$ and $17.19$ (Table~\ref{tab:E}). For $k$ even the minimum is attained at $p=0$ and the energy grows with $|p|$, so that a net winding has an elastic cost; for $k$ odd the selected sector is the odd one for $k=1$ and the even one for $k\ge3$. The complete two-dimensional problem, solved by finite elements and checked against a spectral discretisation, an independent implementation and a multistart exploration of $60$ initial conditions per sector, confirms this picture up to $\varepsilon=0.9$. For $k\ge3$ the gap between the two lowest sectors decreases from about $4.3$ to $2.2$--$2.5$ as $\varepsilon$ grows, whereas for $k=2$ it is non-monotonic, because the minimiser of the sector $p=1$ changes branch at $\varepsilon\simeq0.60$. For $k=1$ the selected sector changes from odd to even at $\varepsilon_c\simeq0.58$; this threshold is displaced by the elastic anisotropy, mainly through the ratio $k_1/k_3$, a stiff splay constant favouring the odd sector and a large twist constant the even one when $k_1/k_3\lesssim1$. The sector selection is therefore robust for $k\ge2$ in the isotropic case and delicate for $k=1$, which is physically the most relevant case. The minimisers depend only weakly on the transverse coordinate, the variation of $\alpha$ across the width never exceeding $8.5^\circ$ for $k\le6$ and $\varepsilon\le0.9$ ($2.4^\circ$ for $k=1$). Finally, the energy is invariant under $(k,u,\alpha)\mapsto(-k,-u,-\alpha)$, so that it does not distinguish between the two enantiomorphic strips: this reflects the achirality of the Frank energy, whose extrinsic part depends on $\Lv$ only through $|\Lv\nv|^2$, and contrasts with the ferromagnetic M\"obius ring of \cite{PK15}, where the exchange energy couples the chirality of the ring to that of the texture.

The limitations of the present analysis are the following. We have computed minimisers, the lowest found from many initial conditions but not proved to be global, together with a few metastable ones; the unstable branches and the bifurcation structure remain to be explored. The director is tangent to the strip and the edge is free; a finite anchoring would introduce a further control parameter and may modify the sector selection, particularly near the crossing for $k=1$. Only smooth line fields are considered: for large $k$ and large aspect ratios textures with defects may have a lower energy, and the minimisers with $k\ge4$ are the lowest smooth states; this regime requires a $\Qv$-tensor formulation.

The most natural extension is the dynamics. The two-dimensional Ericksen--Leslie equations for an active nematic on a surface, as in \cite{NT20}, can be formulated on the orientable double cover (a device that the equilibrium problem studied here does not require), on which the vorticity and the stream function change sign across the identification edge, and the non-uniform equilibrium branches determined here provide the base states for the corresponding stability analysis. The results above suggest questions that are specific to this geometry: whether the topological sector $p$, which for $k$ even cannot change without the nucleation of defects, is preserved by the flow; how the metastable sectors and the crossing at $\varepsilon_c$ for $k=1$ affect the response to a slow change of the aspect ratio; and what flows an active stress drives on a M\"obius strip. A further direction is the extension to $n$-atic order, in which the order parameter is invariant under rotations by $2\pi/n$ and is described by an angle $\theta$ defined modulo $2\pi/n$; the ferromagnetic ring of \cite{PK15} corresponds to $n=1$ and the nematic case treated here to $n=2$. Since the reversal of $\ev_u$ at the identification edge for $k$ odd maps $\theta$ into $\pi-\theta$, the identification condition reads $\theta(u,0)=-\theta(-u,2\pi)+\pi+2\pi p/n$ for $k$ odd and $\theta(u,0)=\theta(u,2\pi)+2\pi p/n$ for $k$ even. The offset $\pi$ is a multiple of $2\pi/n$ if and only if $n$ is even: in that case it is absorbed in the label $p$ and the sector structure of Sect.~III is recovered, whereas for $n$ odd it cannot be absorbed, so that the sectors of the odd strips are of a different kind. The intrinsic part of the energy extends directly, but the extrinsic coupling has to be specified for each $n$, and we have not attempted it. We hope that the combination of topology, extrinsic curvature and nematic order in a non-orientable geometry will stimulate further work, including other non-orientable surfaces, and the experimental observation of the predicted sector crossing.

\section*{Acknowledgments}
The work of G.~Napoli is carried out under the auspices of GNFM--INdAM.

\appendix
\section{Numerical solution of the reduced problem}\label{app:red}
We minimised $\mathcal E$ for $k=0,\dots,5$ and $p=0,\dots,3$. For $k$ even $\alpha-(-pv/2)$ was expanded in a Fourier series with integer modes, for $k$ odd $\alpha-p\pi/2$ was expanded in half-integer modes, which implements the antiperiodicity in \eqref{bc21}; the minimisation was performed by a quasi-Newton method started from $120$ random initial conditions, with $14$ harmonics. Increasing the number of harmonics from $8$ to $14$ changes the energies by less than $2\times10^{-3}$. The lowest energies found are collected in Table~\ref{tab:E} and Fig.~\ref{fig:red}. These results are the best of a finite number of initial conditions and do not exclude the existence of lower minima. The minima of Table~\ref{tab:E} for $k=1,\dots,4$ and $p=0,\dots,3$ were checked with an independent method: \eqref{ode} with the conditions \eqref{bc} was solved as a boundary value problem (\texttt{bvp4c} of MATLAB) starting from random initial guesses and from the minimisers above, and the energy \eqref{Ered} was evaluated a posteriori on each solution. The two methods agree to within $2\times10^{-7}$ (to within $8\times10^{-6}$ for $k=4$, $p=3$, which we attribute to the truncation of the Fourier series), and the equivalence $p\sim p+2$ for $k$ odd is reproduced. The boundary value problem also yields unstable solutions, with a Morse index of $1$ or $2$, and shows that further local minima of higher energy exist in some sectors (for instance $E\simeq7.68$ for $k=1$, $p$ odd, and $E\simeq20.21$ for $k=3$, $p$ even), which the minimisation from random starts had not found in those sectors; the values of Table~\ref{tab:E} for $k=0$ and $k=5$ have not been checked in this way.

\section{Discretisation and validation of the two-dimensional problem}\label{app:2D}
\subsection{Discretisation}
The unknown $\alpha$ is represented by bilinear ($Q_1$) finite elements on a uniform mesh of $N_u\times N_v$ cells in $(u,v)\in[-\varepsilon,\varepsilon]\times[0,2\pi)$, with $2\times2$ Gauss quadrature; the identification edge is imposed exactly by adding a column of nodes at $v=2\pi$ that is expressed through the nodes at $v=0$ by \eqref{bc} (including the reflection $u\to-u$ and the shift $p\pi$), and the free boundary condition \eqref{bc:free} (or \eqref{bc:aniso}) is the natural condition of the discrete functional. The discrete energy and its exact gradient are minimised by a limited-memory quasi-Newton method (L-BFGS-B~\cite{BLNZ95}; among quasi-Newton methods it was chosen because it needs only the energy and its exact gradient, because its memory and cost per iteration grow linearly with the number of unknowns ($\simeq2\times10^3$ here), so that the Hessian is never assembled, and because, being a descent method with line search, it is attracted to local minima and not to saddle points, unlike Newton's method; the bound constraints that distinguish the \emph{B} variant are not used); each sector $(k,p)$ is explored by starting from the reduced minimiser of Sect.~IV, continued in $\varepsilon$ from $\varepsilon=0.05$ to $0.9$, and by random initial conditions of three types (smooth perturbations of the reduced solution with a few low Fourier modes, nodal fields uniformly distributed in $[-\pi,\pi]$, and nodal Gaussian fields). The results below were obtained with $N_u=16$, $N_v=128$.

\subsection{Convergence and independent checks}
(a) For $\varepsilon=0.05$ the numerical energies, $\hat{\mathcal E}=2.116$ ($k=1$, $p$ even) and $1.519$ ($k=1$, $p$ odd), agree with the reduced values of Table~\ref{tab:E} ($2.11$ and $1.51$). (b) Refining the mesh from $16\times128$ to $32\times256$ changes $\hat{\mathcal E}$ by less than $10^{-3}$ (relative $10^{-4}$): for $\varepsilon=0.5$ we find $2.1104\to2.1102$ ($k=1$, $p$ odd), $2.2935\to2.2935$ ($k=1$, $p$ even) and $11.331\to11.330$ ($k=3$, $p$ even). An independent spectral discretisation of the same functional (Chebyshev polynomials in $u$, Fourier series in $v$, identification conditions built into the basis) gives the converged values $2.11015$, $2.29347$ and $11.33005$, so that the mesh error of the $16\times128$ results is $2.5\times10^{-4}$, $6\times10^{-5}$ and $1.3\times10^{-3}$; at $\varepsilon=0.9$ and $k\le3$ the largest change between the two meshes is $1.2\times10^{-3}$, and the order of the sectors is unchanged. Gradient quantities converge more slowly: $\max|\grads\nv|$ changes by at most $1.7\%$ and the variation of $\alpha$ across the width by at most $0.06^\circ$. The spectral code reproduces the crossing for $k=1$ at $\varepsilon_c=0.576$ and the values quoted in Sects.~V.D and V.E, and its minimisers satisfy the natural boundary conditions \eqref{bc:free} and \eqref{bc:aniso} with residuals that decrease as the number of modes grows ($10^{-7}$ for equal constants, $2\times10^{-3}$ for $k_1/k_3=1/4$, $k_2/k_3=4$ with the largest basis).

(c) We repeated the minimisation from $60$ random initial conditions per sector, $20$ for each of the three types listed above, for $k=1,2,3$ (all the sectors) at $\varepsilon=0.3,\,0.6,\,0.9$. The lowest minimum found was reached by starts of each type, and all the minima found are true minima of the discrete energy (the Hessian, computed by finite differences of the exact gradient, is positive definite). In all cases but one the lowest minimum is that of the branch obtained by continuation from the reduced solution; the exception is $k=2$, $p=1$ for $\varepsilon\gtrsim0.60$, where two minima exchange their order (Sect.~V.B). Further local minima, such as $4.15$ against $2.17$ for $k=1$, $p$ even, $\varepsilon=0.3$, are reached only from fields with a localised wall of height $\pi$ in the director angle; minima of this kind exist in $8$ of the $15$ sectors with $k\le6$. The metastable branches are discarded below, and the minima reported are the lowest found, not proved to be global.

(d) For random fields $\alpha(u,v)$ and $k=1,\dots,4$, the quantities $\kappa_t^2$, $\tau_n^2$ and $\kappa_n^2+c_n^2$ of Sect.~II coincide, to machine precision, with the squares of splay, twist and bend computed from the three-dimensional director field extended independently of $\Zt$, which confirms the form of the density, including the role of $k_2$. (e) A further implementation of the $Q_1$ discretisation, written from \eqref{Ehat} and Sect.~II only, with its own limited-memory quasi-Newton minimiser (memory $20$, tolerance $10^{-9}$ on the gradient; the default tolerances of SciPy's L-BFGS-B change the energies by less than $6\times10^{-7}$), reproduces the energies of Table~\ref{tab:k1} at $\varepsilon=0.5,\,0.6,\,0.9$, the energy differences of Fig.~\ref{fig:aniso} at the three points at $\varepsilon=0.3$ of Sect.~V.E ($-0.645$, $+0.288$ and $-0.473$ at the isotropic point) and the mirror symmetry of Sect.~V.F. With it the minima of all the sectors with $k=1,\dots,6$ were recomputed for the twelve values of $\varepsilon$ (ten of them shown in Figs.~\ref{fig:eps} and \ref{fig:phase}, which omit $\varepsilon=0.55$ and $0.65$), from the minima at the previous $\varepsilon$, from six random fields (two of each type) and from two fields with a wall of height $\pi$; the data of those figures are the result.

\end{document}